\documentclass[%
 reprint,
 amsmath,amssymb,
 aps,
]{revtex4-2}

\usepackage{graphicx}
\usepackage{dcolumn}
\usepackage{bm}
\usepackage{lineno}
\usepackage{float}
\usepackage[utf8]{inputenc}
\usepackage[T1]{fontenc}
\usepackage{booktabs, array, mathptmx, float, tabularx, booktabs, lipsum, amsmath,multirow}
\usepackage{siunitx, xcolor}
\usepackage[version=4]{mhchem}
\graphicspath{{figs/}{figsgaoerb/}} 
\usepackage[colorlinks,linkcolor=blue,anchorcolor=blue,citecolor=blue]{hyperref}
\usepackage{braket}
\usepackage{footnote}
\usepackage{mathrsfs}
\usepackage{amsmath}
\usepackage{amssymb}
\usepackage{wasysym}
\usepackage{color,soul}
\usepackage{physics}
\usepackage{siunitx}
\usepackage{dsfont}
\usepackage{float}
\usepackage{lineno}
\usepackage[english]{babel}
\usepackage{blindtext}
\usepackage[english,nomargin,inline,marginclue,draft]{fixme}
\pdfpageheight\paperheight
\pdfpagewidth\paperwidth
        
\fxusetheme{colorsig}
\FXRegisterAuthor{cg}{acg}{CG}  
\FXRegisterAuthor{th}{ath}{\color{blue}TH}  
\FXRegisterAuthor{ib}{aib}{\color{red}IB} 
\FXRegisterAuthor{sh}{ash}{\color{cyan}SH} 
\FXRegisterAuthor{db}{adb}{\color{green}DB} 
\FXRegisterAuthor{ps}{aps}{PS}
\makeatletter
\renewcommand*\FXLayoutInline[3]{%
  {\@fxuseface{inline}\ignorespaces{\color{fx#1}[#3: #2]}}}
\makeatother

\long\def\symbolfootnote[#1]#2{\begingroup%
\def\thefootnote{\fnsymbol{footnote}}\footnotetext[#1]{#2}\endgroup}

\def\nobreakbefore{%
  \relax\ifvmode\else
    \ifhmode
      \ifdim\lastskip > 0pt\relax
        \unskip\nobreakspace
      \else 
        \nobreakspace
      \fi
    \fi
  \fi
}
\let\oldcite\cite
\renewcommand\cite{\nobreakbefore\oldcite}
\begin{document}

\title{On-Chip Quantum States Generation by Incoherent Light}

\author{Yue-Wei Song$^{1,2}$}

\thanks{Y.W.S, H.Z and L.C contribute equally to this work.}

\author{Heng Zhao$^{1,2,4}$}

\thanks{Y.W.S, H.Z and L.C contribute equally to this work.}

\author{Li Chen$^{1,2}$}

\thanks{Y.W.S, H.Z and L.C contribute equally to this work.}

\author{Yin-Hai Li$^{1,2}$}
\author{En-Ze Li$^{1,2}$}
\author{Ming-Yuan Gao$^{1,2}$}
\author{Ren-Hui Chen$^{1,2}$}
\author{Zhao-Qi-Zhi Han$^{1,2}$}
\author{Meng-Yu Xie$^{1,2}$}
\author{Guang-Can Guo$^{1,2,3}$}
\author{Zhi-Yuan Zhou$^{1,2,3}$}
\altaffiliation {Corresponding author: zyzhouphy@ustc.edu.cn}
\author{Bao-Sen Shi$^{1,2,3}$}%
\email{Corresponding author: drshi@ustc.edu.cn}

\address{{$^1$}CAS Key Laboratory of Quantum Information, University of Science and Technology of China, Hefei, Anhui 230026, China\\
{$^2$}CAS Center for Excellence in Quantum Information and Quantum Physics, University of Science and Technology of China, Hefei 230026, China\\
{$^3$}Hefei National Laboratory, University of Science and Technology of China, Hefei 230088, China\\
{$^4$}Chongqing United Microelectronics Center Co. Ltd, Chongqing 400000, China}





\begin{abstract}
On-chip quantum sources based on nonlinear processes are pivotal components in integrated photonics, driving significant advancements in quantum information technologies over recent decades. Usually, the pump coherence has been considered to be crucial for ensuring the quality of generated states, therefore incoherent light is rarely used in quantum information processing. In this work, we explore and reveal the constructive influence of pumped temporal incoherence on the quantum properties of photon sources. Taking silicon waveguides as nonlinear media, we theoretically show that temporal incoherence of light can improve pumping utilization efficiency, resulting in higher source brightness in a spontaneous four-wave mixing process, and the spectrally uncorrelated nature of incoherent light is transferred to the generated photon source, allowing high-purity state preparation. Experimentally, we obtain a higher photon pair generation rate and the lower heralded second-order autocorrelation with an Amplified Spontaneous Emission source. Additionally, we successfully generate a polarization-entangled state with Bell inequality violation of $S$ = 2.64 ± 0.02 and a fidelity of 95.7\% ± 0.1\%. Our study reveals the mechanism behind incoherently pumped quantum states and presents a method for generating photon sources using an easily accessible incoherent light.

\end{abstract}

\maketitle


For decades, integrated photonics has played a transformative role in advancing the development and exploration of the quantum information processing. It is not only essential for ensuring the scalability and compactness of quantum systems, but also for deepening our understanding of fundamental quantum principles\cite{1,2,3,4,5,6,7}. The evolution of on-chip systems has been driven by the pursuit of more advanced or practical solutions, evolving from basic straight waveguides to all-encompassing platforms\cite{8,9,10,11}. Among the distinct configurations within quantum circuits, an integrated photon-pair source stands as a crucial element\cite{12,13,14,14.1,14.2}. Owing to its pivotal role in quantum systems\cite{15,16,17,18,19,20,21}, it is indispensable for quantum information processing\cite{22,24,25}. 

In the integrated systems, quantum photon sources are typically generated via parametric processes. There have been various photonic platforms demonstrating intrinsic advantages in previous studies. Among them, silicon devices are the most widely used due to their high third-order susceptibility and the compatibility with mature complementary metal-oxide semiconductor (CMOS) fabrication techniques\cite{26,27,28,29}. They have been the foundation for many exploratory works requiring intricate structures\cite{n1,n2,n3}.

A prevailing belief in quantum state generation is that the coherence of the pumping dictates the properties of the resulting sources. Incoherent light is predominantly utilized in fields such as imaging, computing, and remote sensing, with limited use in quantum information processing\cite{n4,n5,n6,43.2,n7,n8}. However, the intrinsic decoherence effect arising from laser linewidth broadening poses a fundamental limitation for coherent light applications in long-distance quantum communication and sensing. This phenomenon leads to a progressive accumulation of quantum noise as the system exceeds the coherence time limit\cite{n9,n10,n11}. For monolithic integrated systems, the fabrication of high-performance on-chip lasers also remains a non-trivial challenge. Currently, there are increasing interests in utilizing more natural light sources, such as incoherent or partially coherent beams, to generate quantum resources\cite{36,37,38,43.1}. Considerable progress has been made in understanding the role of coherence in Spontaneous Parametric Down-Conversion (SPDC), with notable achievements in both the theoretical derivation and experimental characterization of quantum states pumped by incoherent light\cite{38,40,41,42,43}. 

Despite these promising developments, the role of temporal incoherence remains underexplored, particularly in integrated systems. Previous studies have focused on $\chi ^{(2)}$ processes in bulk materials, where strict phase-matching conditions inherently constrain the investigation of frequency-domain incoherence\cite{n12,36}. Additionally, the quantum correlations of generated sources also have not been systematically characterized. This gap not only hinders the optimization of quantum light sources, but also obscures the fundamental physics of temporal incoherence and quantum properties in nonlinear processes.

To address the issue, using an Amplified Spontaneous Emission (ASE) source as pumping for photon-pair generation via on-chip Spontaneous Four-Wave Mixing (SFWM) process presents a promising approach. ASE source has high spatial coherence and low temporal coherence\cite{n13,n14,43.2}. In SFWM, the generated photons exhibits a continuous broadband emission spectrum (compared with filtered ASE spectrum), and the pumping bandwidth does not limit the conversion efficiency of the parametric process. Meanwhile, the Gaussian spatial mode allows for low-loss propagation of light within the waveguide.

In this work, we reveal a counterintuitive phenomenon where temporal incoherence plays a beneficial role in quantum state generation. Based on the SFWM process in silicon waveguides, we demonstrate the enhanced photon source preparation by incoherent light, challenging the conventional paradigm that relies on coherent laser. Temporal incoherence can improve pump utilization in nonlinear process. The spectral uncorrelation is imprinted from ASE source to the correlated photon pairs, which enhances the purity of state. We analyze the effects theoretically and predict the results based on analytical expressions and a numerical model.
Beyond silicon waveguides, our theoretical framework is universally applicable to SFWM processes in alternative photonic structures, such as bulk $\chi^{(3)}$ nonlinear crystals, demonstrating broad adaptability across material systems.

Experimentally, we characterize the correlation and entanglement properties of states pumped by filtered ASE source (incoherent) and continuous-wave (CW) lasers (coherent). In addition to over 40\% higher photon generation rate (PGR), incoherently pumped source shows noticeable advantages in the coincidence-to-accidental ratio (CAR) and heralded second-order autocorrelation $g_{H}^{( 2)}( \tau )$ at low power. For the prepared polarization entanglement, we achieve Bell value $S$ = 2.64 ± 0.02 and fidelity of 95.7\% ± 0.1\% for incoherently pumped state. By demonstrating a novel approach to generate high-quality quantum states using incoherent light, this work systematically establishes a framework for bridging incoherent photonic excitation with quantum state preparing. Our research not only relaxes the technical constraints on laser sources, but also paves the way for monolithic quantum photonic integration.

\section*{Coherently and incoherently pumped $\chi^{(3)}$ processes}
In this study, we investigate the generation of a two-photon state via SFWM process. Assuming the pumping power is sufficiently low to avoid two-photon absorption and multiphoton events, the process is governed solely by the following Hamiltonian:
\begin{equation}
H=\frac{3}{4} \epsilon _{0} \chi ^{( 3)}\int d^{3}\boldsymbol{r}\hat{E}_{1}^{( +)}\mathnormal{(}\boldsymbol{r}\mathnormal{,t)}\hat{E}_{2}^{( +)}\mathnormal{(}\boldsymbol{r}\mathnormal{,t)} \cdot \hat{E}_{s}^{( -)}\mathnormal{(}\boldsymbol{r}\mathnormal{,t)}\hat{E}_{i}^{( -)}\mathnormal{(}\boldsymbol{r}\mathnormal{,t)},
\end{equation}
where $\epsilon_0$ and $\chi^{(3)}$ are the vacuum permittivity and third-order susceptibility. The spatial integrals represent the light propagation in the medium, with the superscripts '+' and '-' in the electric field operator corresponding to the annihilation of pumping fields and the creation of correlation properties, respectively. Due to the intensities of the generated and pumping fields, the electric fields can be expressed as follows:
\begin{eqnarray}
\hat {\boldsymbol{E}}^{( -)}(\boldsymbol{r},t) &&=\mathnormal{f^{*}( x,y})\int \sqrt{\frac{\hbar \omega }{2\epsilon _{0} n( \omega ) c}} a^{\dagger }( r,t)\frac{e^{-i[ k( \omega ) z-\omega t]}}{\sqrt{2\pi }} d\omega ,\nonumber\\
\boldsymbol E_{C}^{( +)}(\boldsymbol{r},t) &&=Af( x,y)\int d\omega \alpha _{C}( \omega )  e^{-i( \omega t-kz)}, \nonumber \\
\boldsymbol E_{I}^{( +)}(\boldsymbol{r},\omega) 
&&=Af( x,y) \alpha _{I}( \omega ) e^{-i\phi ( \omega )} e^{-i( \omega t-kz)}  ,
\end{eqnarray}
where $a^{\dagger }(r,t)$ is the creation operator, $n$ is the refractive index as a function of frequency, $A$ is a constant, and $f(x,y)$ is the normalized transverse spatial distribution describing the profile of the guided mode. The pumping field $\boldsymbol E_{C,I}^{( +)}$ is expressed in classical form for the high intensity compared to the states excited from vacuum fluctuations, and the subscript represents the coherence of the optical field.

As a temporally highly incoherent source, ASE is generated without a resonant cavity, resulting in a broadband, smoothed, and stable spectrum. Compared with coherent laser, an additional phase factor $\phi(\omega)$, randomly distributed in the range [0,$2\pi$], illustrates the temporal incoherence of the ASE field. Due to the lack of correlation between each component in frequency, the spectral distribution $\alpha(\omega)$ carries different implications. For coherent light, it represents the normalized amplitude density function, where $\int |\alpha_{C}(\omega)|^2 d\omega = 1$. However, uncorrelated independent components lead to the result $< E(\omega_i) * E(\omega_j) > = 0, i \neq j$. The electric field of ASE illumination needs to be expressed in discrete form. This distinction is also reflected in the subsequent photon pair generation process. In this scenario, the power of coherent and incoherent light are respectively given by
\begin{eqnarray}
P_{C} &&=\frac{1}{2} n\epsilon _{0} cA^{2}{\left| \int \alpha_{C}( \omega ) d\omega \right| ^{2}}, \nonumber\\
P_{I} &&=\frac{1}{2} \epsilon _{0} ncA^{2} \Sigma _{n=1}^{\infty }| \alpha _{I}( \omega _{n})| ^{2}.
\end{eqnarray}

By applying first-order perturbation theory, the two-photon state can be written as \cite{44,45,46}.
\begin{equation}
\ket{\psi } =\ket{0} +\iint F( \omega _{s} ,\omega _{i}) I( \omega _{s} ,\omega _{i}) a^{\dagger }( \omega _{s}) a^{\dagger }( \omega _{i})\ket{0} d\omega _{s} d\omega _{i},
\end{equation}
where $I(\omega_s, \omega_i) = i\gamma L \text{sinc}(\Delta k L/2) e^{-i\Phi(\omega_s, \omega_i)}$ represents the nonlinear effects in the waveguide and the phase-matching conditions. It is determined by the pumping frequency and the detuning frequency. Spectral broadening of the coherent pumping enables the generation of photon pairs that are not symmetric about the central wavelength. $\omega_{0}$, which is illustrated by the two photon amplitude $F( \omega _{s} ,\omega _{i}) =\xi_{C} \int d\omega \mathnormal{_{p}} \alpha_{C}( \omega _{p}) \alpha_{C}( \omega _{i}+\omega _{s} -\omega _{p})$. For coherent laser, $\xi_{C} = \frac{1}{2} \epsilon_0 n c A^2 \int \alpha_C(\omega) d\omega$ is a constant determined by the line-shape and power. The integral term represents the self-convolution of the pumping field, reflecting the overlap of the optical field with itself when the generated photons are asymmetric around $\omega_{0}$. In order to provide an analytical solution, we assume that the coherent pumping field satisfies Gaussian distribution centered at $\omega_0$ with bandwidth $\sigma_p$, and its normalized form is given by $\alpha_{C}( \omega ) =C \cdot exp\left\{-\frac{( \omega -\omega _{0})^{2}}{2\sigma _{p}^{2}}\right\}$. Integrating over the given range frequency yields the production rate of photon pairs. We assume symmetric distribution of signal and idler photons about $\omega_0$, with a detuning unit interval $\Delta \Omega$ significantly larger than $\sigma_p$. The detuning range is defined as $\Omega(m) = m * \Delta \Omega$ ($\omega _{i,s} =\omega _{0} \pm \Omega (m)$), where $m$ represents the number of intervals and $\Delta \Omega$ is the frequency shift per interval. In SFWM, the phase-matching function varies slowly with frequency, so within a detuning interval, $I(\omega_s, \omega_i)$ can be treated as constant. Therefore, the photon generation rate $\Delta N_{C}(\Omega(m))$ is given by:
\begin{equation}
\Delta N_{C}( \Omega (m)) = \frac{\Delta \Omega }{2\pi } \gamma ^{2} L^{2} P_{C}^{2} sinc^{2}\left(\frac{\Delta k_{m} L}{2}\right)\frac{\sqrt{2}}{2}.
\end{equation}
The entire generating rate can be obtained through direct summation. See more details in Supplementary Information. For incoherent pumping, the characteristic of each component being mutually uncorrelated alters the interaction mode for SFWM. The inner product of the two-photon amplitude $\mathcal{F}_{I}$ can be expressed as
\begin{equation}
\mathcal{F}_{I} =\mathnormal{\left| \xi_I\sum _{p=1}^{\infty } \alpha_I ( \omega _{p}) e^{-i\phi ( \omega \mathnormal{_{p}})} \alpha_I ( \omega _{s} +\omega _{i} -\omega _{p}) e^{-i\phi ( \omega \mathnormal{_{i+s-p}})}\right| }^{2}.
\end{equation}

Unlike the case where the power is calculated by summing up the amplitudes of coherent light, the presence of $\phi(\omega)$ characterizing the incoherence ensures independent interactions between the each two components. Even for the photons generated in same mode, the contributions of different combinations $A_{m} =\alpha_{I} \mathnormal{( \omega _{m}) \alpha_{I} ( \omega _{l} +\omega _{k} -\omega _{m})}$ are still irrelevant, i.e.
\begin{equation}
\left< \left( \Sigma A_{m} e^{-i\phi _{m}}\right)^{2}\right> =\Sigma \left(\left| A_{m} e^{-i\phi _{m}}\right| ^{2}\right).
\end{equation}
Based on this relationship, we apply the same transformation as in Eq 5, the PGR for incoherent pumping is denoted as
\begin{equation}
\Delta N_{I}( \Omega (m)) =\frac{\Delta \Omega }{2\pi } \gamma ^{2} L^{2} P_{I}^{2} sinc^{2}\left(\frac{\Delta k_{m} L}{2}\right).
\end{equation}
See more deduction details in Supplementary Information. This conclusion is consistent with the expression for monochromatic pumping\cite{44}, as both cases share the same underlying physical interpretation. For coherent lasers with spectral broadening, when generating photon pairs with an asymmetric distribution relative to the center wavelength, the energy conservation condition limits the pumping utilization efficiency. This is reflected in the optical field's overlap, represented by the self-convolution. ASE light can be viewed as a linear combination of $N$ independent monochromatic light sources. These components interact in the medium to generate photon pairs. The total arrangement ($A_{N}^{n} =\frac{N!}{( N-n) !}$) is given by $A_{N}^{1} +A_{N}^{2}=N^2$, where $A_{N}^{1}$ corresponds to degenerate pumping, and $A_{N}^{2}$ represents the combinations of non-degenerate pumping. Their sum equals to $\sum _{i=1}^{\infty }P_{i} \sum _{j=1}^{\infty } P_{j} =P^{2}$. 

The impact of optical field coherence is also evident in the state purity. For a non-degenerate process, the joint spectral amplitude (JSA) of photons generated in SFWM results from the interaction of different frequency components, which superpose in the same mode $\ket{1_{\omega _{s}} ,1_{\omega _{i}}}$. Coherent light fields exhibit the characteristic of amplitude combination due to phase consistency, while incoherent light fields demonstrate intensity summation with no correlation between them. The interacting modes of both can be represented as follows:
\begin{eqnarray}
f_{C}( \omega _{s} ,\omega _{i}) &&\propto \int E_{C}( \omega _{p}) *E_{C}( \omega _{s} +\omega _{i} -\omega _{p}) d\omega_p \nonumber \\
f_{I}( \omega _{s} ,\omega _{i}) &&\propto \sum _{i=1}^{\infty } E_{I}( \omega _{p}) *E_{I}( \omega _{s} +\omega _{i} -\omega _{p}) e^{i\phi ( \omega_p )}.
\end{eqnarray}

In numerical simulations, the former approach corresponds to directly summing the terms, while the latter involves summing the squared terms and then taking the square root of the sum. For coherent light, the energy tends to concentrate symmetrically around the center wavelength of the pumping field. In contrast, due to the absence of phase correlation, incoherent light shows a more dispersed energy distribution, which achieves frequency-uncorrelation. 
\begin{figure}[h]
\centering
\includegraphics[width=0.49\textwidth]{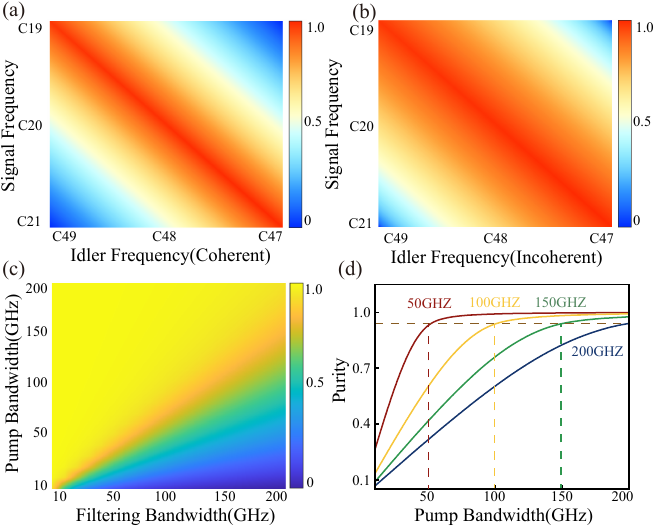}
\caption{\label{fig:epsart}Joint Spectral Amplitude and Purity (a) Joint spectral amplitude with coherent pumping mode; (b) Joint spectral amplitude with incoherent pumping mode;
(c) Purity as a function of filtering and pumping bandwidth;
(d) Variation of the Purity with four bandwidths.}
\end{figure}

As shown in Figure 1, we calculate the JSA and purity for different pumping coherence conditions, using a filtering bandwidth and pumping bandwidth of 200 GHz with a rectangular line shape. For the sake of clarity, the following wavelengths will be defined according to the ITU standard. The center frequency is set at C34, with filter centers chosen at C20 and C48. The JSA for incoherent mode shows a more uniform distribution, achieving a purity of 0.94. In contrast, the purity for coherent mode is only 0.84. Considering that the bandwidth of the incoherent light is generally much broader, this discrepancy becomes more pronounced in the straight waveguide.

Figure 1(c) and 1(d) show the influence of incoherent pumping bandwidth and filtering bandwidth. The state purity shows a clear increasing trend with the increase in pumping bandwidth. However, when the pumping bandwidth exceeds the filter bandwidth, the rate of increase in purity significantly slows down. In this case, more correlated photons are detected in the asymmetric channels.

\begin{figure*}[htbp]
\centering
\includegraphics[width=0.85\textwidth]{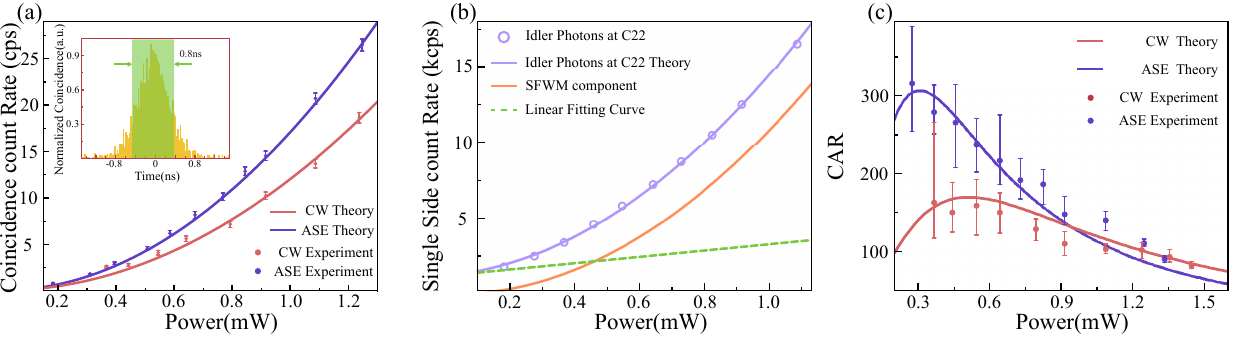}
\caption{\label{fig:wide}
(a) Coincidence count rate versus pumping power;  The insert figure is coincidence histogram of the signal and idler photons at C22 and C46.
(b) Single side count rate of the idler photon from C22 versus pumping power;
(c) CARs versus pumping power. The lines are theoretically calculated by coincidence and single side count.
The error bars of the CAR are obtained by Poissonian photon-counting statistics of coincidence count and background noise fluctuation.}
\end{figure*}

\begin{figure*}[hbt]
\centering
\includegraphics[width=0.85\textwidth]{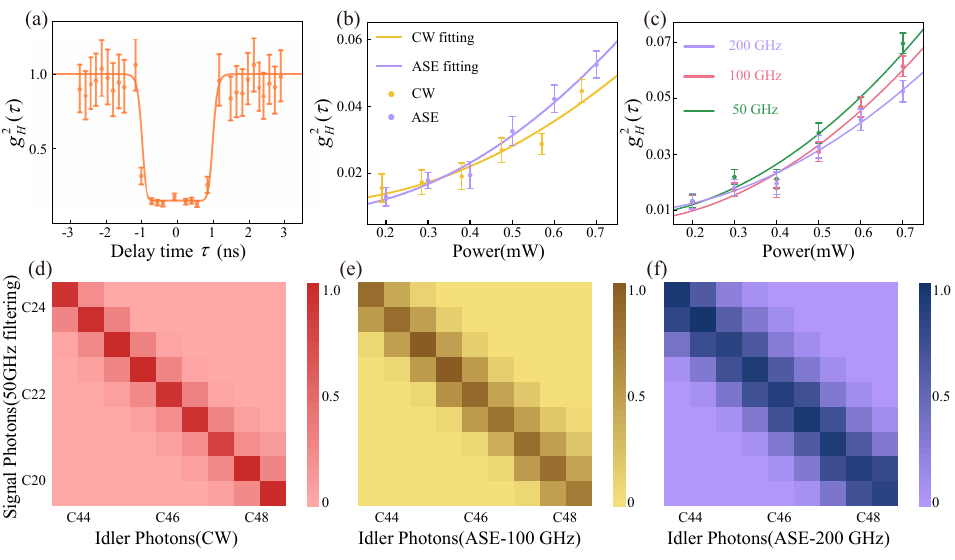}
\caption{\label{fig:epsart}(a) Heralded second-order autocorrelation $g_{H}^{( 2)}( \tau)$ of heralded photon at C20; 
(b) Coherently and Incoherently pumped heralded second-order autocorrelation $g_{H}^{( 2)}( 0 )$ versus pumping power;
(c) Incoherently pumped heralded second-order autocorrelation $g_{H}^{( 2)}( 0 )$ with different bandwidth;
(d) Joint spectral intensity of coherent pumping;
(e) Joint spectral intensity of incoherent pumping with 100 GHZ bandwidth.
(f) Joint spectral intensity of incoherent pumping with 200 GHZ bandwidth.}
\end{figure*}

In "Methods" and Supplementary Information, we develop a numerical model to statistically analyze outcomes, such as coincidence counts and single side counts, as a function of bandwidth or center frequency detuning. These results provide a guidance for selecting the bandwidth of the incoherent pumping to simultaneously ensure both the brightness and purity of the source.

\section*{Results}
\subsection*{Measurements of quantum correlation properties}

\begin{figure*}[htbp]
    \centering
    \includegraphics[width=\textwidth]{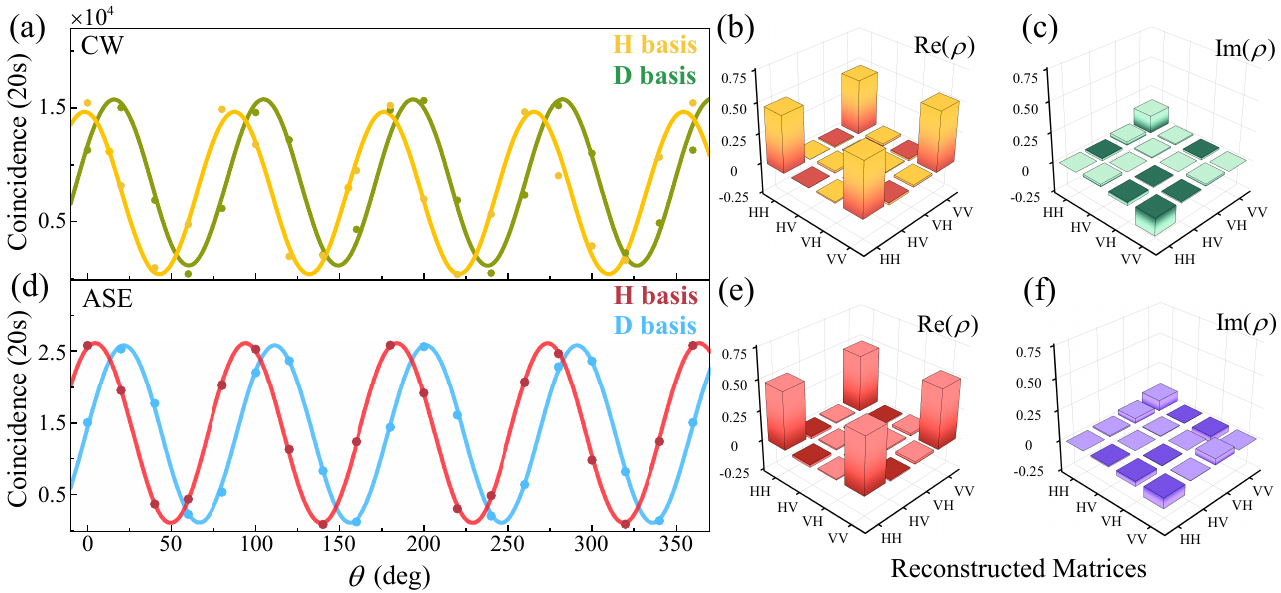}  
    \caption{(a)-(c) Characterization of Polarization Entanglement with coherent pumping. 
(a) Two-photon polarization interference fringes under H and D bases; 
(b),(c) Experimentally measured density matrices;
(d)-(f) Characterization of Polarization Entanglement with incoherent pumping.
(d) Two-photon polarization interference fringes under H and D bases;
(e),(f) Experimentally measured density matrices.}
    \label{fig:banner}
\end{figure*}

In the experiment, we record the single-side count rate and coincidence count rate. Figure 2(a)–(c) present the quantitative characteristics of quantum correlation properties for photon pairs under CW laser and 200 GHz ASE pumping. Figure 2(a) shows the coincidence counts of signal and idler photons at C22 and C46, which closely align with numerical predictions (solid line). Since coincidence counts scale quadratically with pumping power, the proportional difference between coherent and incoherent pumping becomes more pronounced at higher power levels. The single-side count, measured to estimate accidental coincidences, can be expressed as $N_{sc} = N_{SFWM} + a \cdot P + N_{0}$, where $N_{SFWM}$ denotes contributions from photon pairs. $a$ represents the intensity of linear noise from Raman scattering in the fiber and ASE sideband photons. $N_{0}$ accounts for background noise such as detector dark counts. At low pumping power, noise photons significantly contribute to the single-side count. Compared to coherent pumping, incoherent pumping generates more photon pairs, achieving a higher peak CAR—up to 345 for 200 GHz ASE, 

whereas the maximum CAR for a CW laser is approximately 160. At higher pumping powers, accidental coincidences dominate the noise count, causing the CAR curve for coherent pumping to fall behind that of incoherent pumping due to its lower efficiency. In addition, we investigate the bandwidth of incoherent pumping and the multiwavelength correlation properties of photon pairs. See Supplementary Information for more details.

The heralded second-order autocorrelation function $g_{H}^{( 2)}( \tau )$ is a key indicator for characterizing the single-photon properties of the source. When heralded signal photons 
at C20 Channel equals to 0.35 MHz, the incoherent pumped antibunching dip of  $g_{H}^{( 2)}( \tau )$ for the heralded single photons is shown in Fig 3(a). The error bars are estimated by Poissonian photon counting statistics. Fig 3(b), 3(c) shows the $g_{H}^{( 2)}( 0 )$ as a function of pumping power. The second-order autocorrelation function is positively correlated with the photon PGR. As the single side counts increase, larger accidental coincidences negatively affect the purity of the heralded photons. In the condition, ASE pumped source exhibits higher brightness at the same power, resulting in a faster growth rate. However, the single side counts $N_1$ and threefold coincidence events $N_{123}$ are primarily dominated by system noise at low power. In this case, a higher PGR leads to a lower second-order autocorrelation. For incoherent pumping, an increase in bandwidth reduces the efficiency of collecting correlated photons. See more details in the Supplementary Information.

Figure 3(d)-(f) show the joint spectral intensity (JSI) for two types of pumpings, with the center pumping frequency set at C34. A 50 GHz dense wavelength division multiplexing(DWDM) is used for dividing frequency-correlated photons and filtering. The broadening of the pumping light, detuning of the pumping center frequency, and overlap between channels result in coincident events being detected in the asymmetric channels. We compare the JSI under different pumping and filtering bandwidth conditions; see more details in the Supplementary Information.

\subsection*{Characterization of polarization entanglement}
In addition to correlation, entanglement is also a pivotal property of quantum states and a key resource in quantum information. We use a Sagnac interferometer loop to prepare polarization-entangled photons (see “Methods” for details). By adjusting the pumping's polarization state, we can generate the maximally entangled Bell state $\ket{\upPhi ^{+}} =\ket{HH}+\ket{VV}$.

We characterize the degree of entanglement using several methods. Figure 4 shows the two-photon polarization interference fringes and the reconstructed experimental density matrix $ \rho _{ex}$.The net fringe visibility for coherent pumping is 99.6\% ± 0.1\% (H basis) and 95.3\% ± 0.2\% (D basis). For incoherent pumping, the net fringe visibility is 97.2\% ± 0.1\% (H basis) and 97.1\% ± 0.1\% (D basis). The estimated net fidelity, calculated as $F=\left[ Tr\left(\sqrt{\sqrt{\rho _{th}} \rho _{ex}\sqrt{\rho _{th}}}\right)\right]^{2}$, equals to 94.5\% ± 0.2\% for coherent pumping and 95.7\% ± 0.1\% for incoherent pumping, where $Tr$ is the trace and $ \rho _{th}$ is the ideal density matrix. 

Furthermore, we measure the S parameters of the Clauser-Horne-Shimony-Holt (CHSH) inequality. For two polarization settings ($\theta _{s}$ = -22.5°, 67.5°, 22.5°, 112.5°; $\theta _{i}$ = -45°, 45°, 0°, 90°), the $S$ values are 2.59 ± 0.03 (coherent) and 2.64 ± 0.02 (incoherent), with no background noise. These results confirm that the generated state is nearly an ideal Bell state, and the temporal coherence of the pumping does not adversely affect polarization entanglement preparation.

\section*{Discussion and Conclusion}
In this study, we report on-chip quantum states generation using both coherent and incoherent pumping in standard silicon nanowires. Contrary to conventional assumptions that temporal incoherence degrades quantum states, our study reveals its unexpected capacity to improve sources performance in integrated systems. The enhancement effect on photon-pair generation has been theoretically analyzed, and numerical simulations demonstrate the transfer of the spectrally uncorrelated property. Experimentally, the photon-pair generation rate with ASE pumping is over 40\% higher than coherently pumped situation. The increased brightness mitigates the impact of background noise at low power, improving quantum correlation properties, including coincidence and $g_{H}^{( 2)}(0)$. Furthermore, we demonstrate high-quality polarization entanglement properties of the photon source. The incoherent pumping not only facilitates polarization entangled state preparation but also, due to its low temporal coherence, eliminates interference effects caused by end-face reflections in the Sagnac loop, enhancing system stability.

In conclusion, our work presents novel quantum resources without stringent coherence constraints. This challenges the traditional view that high-quality photon sources are only achievable with coherent pumping.Remarkably, the interplay between optical incoherence and quantum properties revealed in this work represents a universal feature of parametric nonlinear processes,making our findings applicable to diverse physical systems where SFWM occurs. The discovery also significantly expands the selectivity of pump sources. ASE, often seen as background noise, can be effectively utilized for photon source generation and entanglement preparation. The enhanced performance and reduced coherence requirements of our approach advance the development of monolithic integrated systems, paving the way for scalable quantum photonics technologies.


\nocite{*}

\bibliography{Article}

\begin{acknowledgments}
This work is supported by the National Key Research and Development Program of China (2022YFB3903102, 2022YFB3607700), National Natural Science Foundation of China (NSFC)(62435018), Innovation Program for Quantum Science and Technology (2021ZD0301100), USTC Research Funds of the Double First-Class Initiative (YD2030002023), and Research Cooperation Fund of SAST, CASC (SAST2022-075).
\end{acknowledgments}

\section*{Author contributions}
Y.W.S., H.Z. and L.C. contributed equally to this work. Z.Y.Z. and B.S.S. conceived the project, supervised the work and acquired the funding. Y.W.S., L.C., H.Z. and Y.H.L. conceived and designed the experiments. Y.W.S. and L.C. developed the theory and carried out the numerical simulations under the supervision of Z.Y.Z. and B.S.S.. M.Y.G., R.H.C., Z.Q.Z.H. and M.Y.X analyzed the data. H.Z. and L.C. fabricated the device. Y.W.S., L.C. and M.Y.G. performed the experiments. Y.W.S., E.Z.L, Z.Y.Z., B.S.S. and G.C.G. wrote the manuscript. All authors contributed to discussions and the interpretation of the results.

\section*{Methods}

\subsection*{Numerical Analysis}
In the section of correlation measurement, the impact of temporal incoherence is primarily reflected in the photon yield. In theory, we consider the pumping to be a narrowband source, resulting in the generated photon pairs being symmetrically counted. However, as a temporally incoherent pumping, the bandwidth $\sigma_A$ of the ASE source is comparable to the DWDM channel bandwidth. It is also difficult to ensure that the central frequency $\omega_0$ of the pumping exactly matches the central frequency $\omega_d$ of the DWDM. Therefore, we need to discretize both the pumping spectrum and DWDM channels using numerical methods for accurate calculation. 
\renewcommand{\figurename}{Extended Data Figure}
\renewcommand{\thefigure}{1}
\begin{figure*}[hbtp]
    \centering
    \includegraphics[width=\textwidth]{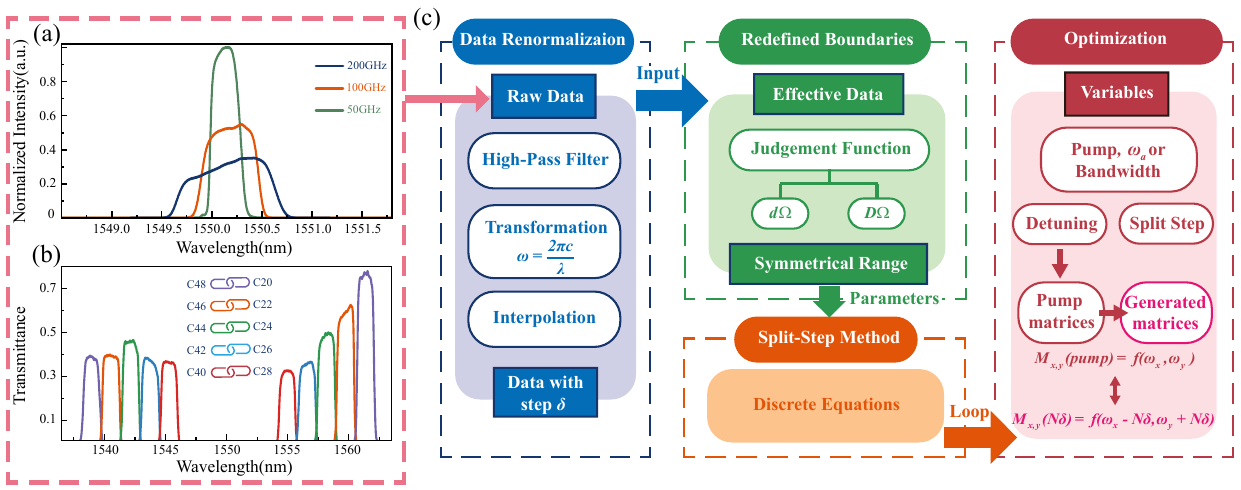}  
    \caption{(a) Normalized input spectrum distribution filtered by 50 GHz, 100 GHz, 200 GHz optical filters; (b) Measured transmittance of 10-channel DWDM; (c) Modeling framework of photon pair generation and transmission in optical system}
    \label{fig:banner}
\end{figure*}
Here, we present a universal model for quantitatively analyzing the collection of broadband photon pairs generated in SFWM. Signal and idler photons are counted separately, reflecting the limitations imposed by the finite collecting bandwidth of DWDM channels on asymmetrically distributed pumping components. In simulations, ideal parameters, such as smoothing rectangular pumpings and lossless DWDM, are used to numerically study the single-side and coincidence counts as a function of bandwidth and detuning frequency. Considering the actual experimental setup and combining the measured spectral line-shape and channel transmittance shown in Figure 1, the calculated results closely match the experimental data. The numerical simulation process is outlined in four steps, as shown in Extended Data Figure Figure 2(c). For detailed steps, refer to the Supplementary Information.

For comparison with experiment results, we should take into account losses during transmission, division, and detection. The actual detectable coincidence count rate can be expressed as $N_{cc} =\mu _{ti} \mu _{ts} \mu _{di} \mu _{ds} N_{0}$, where $N_0$ is the number of generated photons, $\mu_t$ and $\mu_d$ represent the transmission and detection efficiencies, respectively. As shown in Extended Data Figure 2(a) and 2(b), the measured parameters are functions of frequency. Interactions between the pumping components lead to the separate generation of idler and signal photons within the discretized DWDM channels.

\subsection*{Experimental Setups}
Extended Data Figure 2 illustrates the experimental setup. In the pumping source, a tunable CW laser (Toptica DL100) operating at 1550.17 nm serves as the coherent seed laser, while an ASE source with a spectrum spanning 1528 nm to 1563 nm provides the incoherent seed laser. To improve amplification efficiency, a pre-filtering procedure is implemented before the Erbium-Doped Fiber Amplifier (EDFA) for the ASE source. Due to the fully random polarization of the incoherent light, the amplified pumping beam then passes through a polarizer to ensure that the pumping propagates with a single polarization in the waveguide. The tunable attenuator (TA) is used to control input power, and four identical band-pass filters (BPF) are set to suppress sideband noise. 

In the “Correlation” section, a polarization controller (PC) is used to optimize the polarization state for maximum coupling efficiency before the light enters the waveguide. The SOI waveguide, a single silicon nanowire with grating couplers, is used to generate photon-pair state in the system. It is 1 cm long with transverse dimensions of 220 nm in height and 450 nm in width, and has a total insertion loss of 20 dB. Residual pumping light is filtered by cascaded band-stop fiters (BSF), achieving a total rejection exceeding 120 dB. The generated photon pairs have a continuous broadband emission spectrum. The idler and signal photons are separated by a 10-channel 200 GHz DWDM and detected by two free-running InGaAs avalanche photon detectors (APD1, APD2, ID220, detection efficiencies 10\%, dead time 5 $\mu s$). Detection signals are processed by a board (PicoQuant TimeHarp 260) to record coincidence events, with a time bin width of 0.8 ns, as shown in Figure 3(a). Further details on system parameters are provided in Supplementary Information.
\renewcommand{\thefigure}{2}
\begin{figure*}[htbp]
    \centering
    \includegraphics[width=\textwidth]{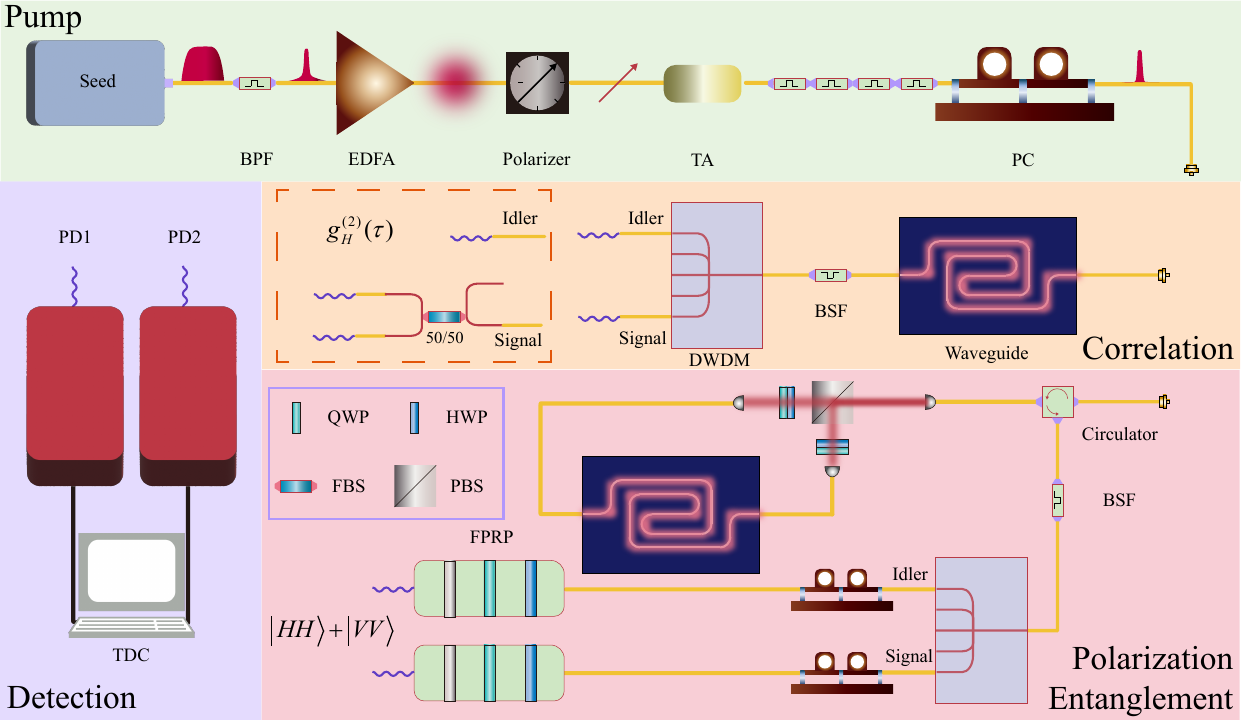}  
    \caption{Schematic diagram of the experimental setup.}
    \label{fig:banner}
\end{figure*}
The heralded second-order autocorrelation function $g_{H}^{( 2)}( \tau )$ involves threefold coincidence events, which imposes requirement on the detection efficiency. Therefore, the heralded signal photons ($N_1$) at C20 Channel are detected using a three-channel SNSPD (detection efficiency 80\%), while the heralding idler photons at C48 channel are detected with a delay time $\tau$ after passing through a 50:50 fiber beam splitter (FBS) ($N_2$, $N_3$). The coincidences ($N_{12}$, $N_{13}$) and threefold coincidence events ($N_{123}$) and measured single side counts are recorded simultaneously by a time-to-digital converter (TDC, UQDevices Logic16). 

In the “Polarization Entanglement” section, we employed a Sagnac interferometer configuration, which is a widely used and self-stabilizing method for generating polarization-entangled states. The PC before entering the Sagnac interferometer modulates the output pumping after passing through a circulator to a 45° linear polarization. 

The loop includes two half-wave plates (HWPs), two quarter-wave plates (QWPs), a polarization beam splitter (PBS), and an approximately 1 cm long end-face coupling silicon nanowire with a 5 dB insertion loss. The PBS splits the pumping into clockwise ($\ket{H}$) and counterclockwise ($\ket{V}$) directions. The HWP and QWP in the Sagnac loop are adjusted to align the polarization of both clockwise and counterclockwise pumping along the horizontal axis (TE polarization) in the silicon waveguide. This configuration allows only the TE polarization mode to propagate along the silicon waveguide on the chip, with no polarization rotation occurring during propagation, as confirmed by three-dimensional finite-difference time-domain simulations.

Therefore, in the clockwise direction, the generated photon pair is in the state $\ket{HH}$. Upon passing through the wave plates, the polarization state is rotated to $\ket{VV}$. Conversely, in the counterclockwise direction, the $\ket{HH}$ polarized state remains unchanged at the PBS. After the photon pairs from both counterpropagating directions recombine at the PBS, the resulting entangled state at the output port of the Sagnac loop can be expressed as.
\begin{equation}
\upPhi =\ket{HH}+\eta e^{i\delta }\ket{VV},
\end{equation}
where the $\eta ^{2}$ represents the ratio of the pumping powers for the V and H polarizations, while $\delta$ denotes the phase difference that arises due to the birefringence experienced by the signal and idler photons in the $\ket{HH}$ and $\ket{VV}$ states. The birefringence here originates from components other than the silicon chip in the setup, as only the TE polarization state interacts with the chip. 

The frequency non-degenerate signal and idler photons are separated by the same 200 GHz DWDM. PCs are used to compensate for the polarization changes introduced during the transmission in fibers. By adjusting the fiber polarization rotator and polarizer (FPRP), we can perform quantum state tomography and measure biphoton interference.

\end{document}


\linenumbers

\title{Supplementary Information for “On-Chip Quantum States Generation by Incoherent Light”}

\author{Yue-Wei Song$^{1,2}$}

\thanks{Y.W.S, H.Z and L.C contribute equally to this work.}

\author{Heng Zhao$^{1,2,4}$}

\thanks{Y.W.S, H.Z and L.C contribute equally to this work.}

\author{Li Chen$^{1,2}$}

\thanks{Y.W.S, H.Z and L.C contribute equally to this work.}

\author{Yin-Hai Li$^{1,2}$}
\author{En-Ze Li$^{1,2}$}
\author{Ming-Yuan Gao$^{1,2}$}
\author{Ren-Hui Chen$^{1,2}$}
\author{Zhao-Qi-Zhi Han$^{1,2}$}
\author{Meng-Yu Xie$^{1,2}$}
\author{Guang-Can Guo$^{1,2,3}$}
\author{Zhi-Yuan Zhou$^{1,2,3}$}
\altaffiliation {Corresponding author: zyzhouphy@ustc.edu.cn}
\author{Bao-Sen Shi$^{1,2,3}$}%
\email{Corresponding author: drshi@ustc.edu.cn}

\address{{$^1$}CAS Key Laboratory of Quantum Information, University of Science and Technology of China, Hefei, Anhui 230026, China\\
{$^2$}CAS Center for Excellence in Quantum Information and Quantum Physics, University of Science and Technology of China, Hefei 230026, China\\
{$^3$}Hefei National Laboratory, University of Science and Technology of China, Hefei 230088, China\\
{$^4$}Chongqing United Microelectronics Center Co. Ltd, Chongqing 400000, China}





\maketitle


\renewcommand{\theequation}{S\arabic{equation}}

\section{\label{sec:level1}Coherence and Incoherence}

For coherent fields, the intensity should account for the coherent superposition of amplitudes in the same mode, resulting in constructive interference. This is reflected in the relationship between electric field intensity and power, where all frequency components of the coherent light are considered as a single mode, and the modulus squared of the integral of the amplitude density over all components gives the overall result. For incoherent light, the independence between components eliminates the need to define an amplitude distribution. Instead, the power spectral density effectively describes ASE light sources with a large bandwidth and continuous spectral distribution. Coherence is directly evident in the contrast between the power expressions of ASE and CW laser in Eq. 3 and the function $\alpha (\omega)$.

Normalized coefficient of the given amplitude distribution function can be solved through  $\int |\alpha_{c}(\omega)|^2 d\omega = 1$. For a Gaussian distribution with the form $\alpha _{c}( \omega ) =C\cdot exp\left\{-\frac{( \omega -\omega _{0})^{2}}{2\sigma _{p}^{2}}\right\}$, which is a common type of spectral broadening, the coefficient C can be computed as $\sqrt{\frac{1}{\sigma _{p}\sqrt{\pi }}}$. Correspondingly, the power of coherent light with Gaussian distribution is represented as
\begin{equation}
P_{C} =\frac{1}{2} n\epsilon _{0} c2\sqrt{\pi } \sigma _{p} A^{2}.
\end{equation}
For ease of direct comparison between photon number and pump power, we extract all coefficient factors from the joint spectral amplitude. Apart from $\xi_C $, $C^2$ contained in the self-convolution of the pump field should be considered. After integrating the dimensionless convolution function, we obtain $F_{0}( \omega _{s} ,\omega _{i})=\sqrt{\pi } \sigma _{p} exp\left[ -\left(\frac{\mathnormal{\omega '_{i}} +\mathnormal{\omega '_{s}}}{2\sigma _{p}}\right)^{2}\right] \  $, and the overall factor can be written as
\begin{eqnarray}
\mathnormal{\xi_C } &&=\frac{1}{2} \epsilon _{0} ncA^{2}\left(\frac{1}{\sqrt{\sigma _{p}}\sqrt[4]{\pi }}\right)^{3}\sqrt{2\pi } \sigma _{p}\sqrt{\pi } \sigma _{p} \nonumber\\
&&=\frac{1}{2} \epsilon _{0} ncA^{2}\sqrt{2\sqrt{\pi }}\sqrt{\sigma _{p}}.
\end{eqnarray}

Next, to calculate the photon generation rate in the given detuning unit interval, we first square the overall factor and exponential term in the equation, then integrate over the range of signal and idler photons. The specific process is as follows: Firstly, $\mathnormal{\xi_C }$ can be directly calculated. Meanwhile, for $\omega \mathnormal{_{p}}$, $\omega \mathnormal{_{i}}$, $\omega \mathnormal{_{s}}$, if they vary little for the slowly varied phase-matching function, $\mathnormal{F( \omega _{s} ,\omega _{i})}$ can be treated as a constant. When integrating the exponential function, we let $\Omega ( n) =n*\Delta \Omega $ and $\omega '=\omega -\omega _{0}$. According to the preset conditions $\Delta \Omega \gg \sigma _{p}$, we have $\frac{\Delta \Omega }{\sigma _{p}} \approx +\infty $. This integral can be solved using the substitution method.
\begin{eqnarray}
\mathcal{F}_{C} &&=\int _{\Omega ( n)}^{\Omega ( n) +\Delta \Omega } d\mathnormal{\omega '_{i}}\int _{\Omega ( -n)}^{\Omega ( -n) -\Delta \Omega } d\mathnormal{\omega '_{s}} exp\left[ -2\left(\frac{\mathnormal{\omega '_{i}} +\mathnormal{\omega '_{s}}}{2\sigma _{p}}\right)^{2}\right]\nonumber \\
&&=\int _{\Omega ( n)}^{\Omega ( n) +\Delta \Omega } d\mathnormal{\omega '_{i}}\int _{-\Omega ( -n)}^{-\Omega ( -n) +\Delta \Omega } d\mathnormal{\omega _{s}^{''}} exp\left[ -2\left(\frac{\mathnormal{\omega '_{i}} -\mathnormal{\omega _{s}^{''}}}{2\sigma _{p}}\right)^{2}\right]\nonumber \\
&&=\int _{n\Delta \Omega }^{( n+1) \Delta \Omega } dX\int _{\omega _{0} -\frac{\Delta \Omega }{2}}^{\omega _{0} +\frac{\Delta \Omega }{2}} dYexp\left[ -2\left(\frac{Y}{\sigma _{p}}\right)^{2}\right]\mathnormal{\left| J\left( \omega '_{i} ,\omega _{s}^{''}\right)\right| }\nonumber \\
&&=\Delta \Omega \sigma _{p}\int _{n}^{n+1} dX^{'} \cdot \int _{-\infty }^{+\infty } dY'exp\left[ -2Y^{\prime 2}\right]\mathnormal{\left| J\left( \omega '_{i} ,\omega _{s}^{''}\right)\right| }\nonumber \\
&&=\sqrt{2\pi } \Delta \Omega \sigma _{p},
\end{eqnarray}
In the equation, the following functional relationships are applied: $\mathnormal{\omega '_{s}} =-\mathnormal{\omega _{s}^{''}}$, $X=\frac{\mathnormal{\omega '_{i}} +\mathnormal{\omega _{s}^{''}}}{2}$, $Y=\frac{\mathnormal{\omega '_{i} -\omega _{s}^{''}}}{2}$, $Y'=\frac{Y}{\sigma _{p}}$, and $\mathnormal{\left| J\left( \omega '_{i} ,\omega _{s}^{''}\right)\right| }$ is the Jacobian factor, which is equal to 2 in this situation. Combining all the results, the expression for the photon count is:
\begin{equation}
N_{C} =\frac{\Delta \Omega}{2\pi } \gamma ^{2} L^{2} P_{C}^{2} sinc^{2}\left(\frac{\Delta k_{n} L}{2}\right)\frac{\sqrt{2\pi } \cdot 2\sqrt{\pi }}{\left( 2\sqrt{\pi }\right)^{2}}.
\end{equation}
Compared with an ideal monochromatic laser, a factor determined by the line-shape reflects the self-convolution of the coherent field. During the generation of photon pairs which are not symmetric about the central frequency $\omega_0$, certain field amplitudes cannot participate in the interaction due to the conservation of energy. The convolution function illustrates the efficiency of the pump light in this process, and under complete symmetry about $\omega_0$, the generation rate is the same as that of a monochromatic laser.

In contrast, the electric field of incoherent light is better regarded as a set of mutually independent discrete variables. For incoherent cases, $\alpha _{I}$ is more appropriately understood as the envelope function of the power spectral density in the form of $| \alpha _{I}( \omega )| ^{2}$. In the experimental system, ASE is generated without a resonant cavity, resulting in an irregular spectrum. Based on the reasons above, defining a normalization condition for a discretely and randomly distributed $\alpha$ seems unnecessary, and the concept of convolution does not apply to incoherent electric field components. 

Incoherent pumping leads to non-degenerate interactions, where the combination of different frequency components results in a linear superposition of photons generated in the same mode $\ket{1_{\omega _{s}} ,1_{\omega _{i}}}$. Since the discrete form is not convenient for describing the continuous distribution  of $\omega _{s}$ and $\omega _{i}$, an integration over the range $\Delta \Omega$ is required to calculate the total photon yield. Therefore, to transform the discrete summation into a continuous integral, The terms with frequency $\omega_{p'} = \omega_s + \omega_i - \omega_p$ should be treated as continuously distributed within the bandwidth $\Delta \omega$. The density is correspondingly represented as $| \alpha _{I}( \omega _{s} +\omega _{i} -\omega _{p})| ^{2}\frac{1}{\Delta \omega }$. Applying the integral transformation in Eq.S3, this expression can be written as:
\begin{eqnarray}
\mathscr{F}_{I} 
&&=\int _{\Omega ( n)}^{\Omega ( n) +\Delta \Omega } d\omega '_{i}\int _{\Omega ( -n)}^{\Omega ( -n) -\Delta \Omega } d\omega '_{s}* \nonumber \\
&&\sum _{p=1}^{\infty }| \alpha _{I}( \omega _{p})| ^{2}| \alpha _{I}( \omega '_{s} +\omega '_{i} -\omega _{p})| ^{2}\frac{1}{\Delta \omega } \nonumber \\
&&=\left| J\left( \omega '_{i} ,\omega_{s}^{''}\right)\right|  \int _{n\Delta \Omega }^{( n+1) \Delta \Omega } dX*\nonumber \\
&&\sum _{p=1}^{\infty }\left(| \alpha _{I}( \omega _{p})| ^{2}\int _{\omega _{0} -\frac{\Delta \Omega }{2}}^{\omega _{0} +\frac{\Delta \Omega }{2}} dY| \alpha _{I}( Y-\omega _{p})| ^{2}\frac{1}{\Delta \omega }\right) \nonumber \\
&&=2\Delta \Omega \sum _{p=1}^{\infty }\left(| \alpha _{I}( \omega _{p})| ^{2}\sum _{m=p}^{\infty }| \alpha _{I}( \omega _{m})| ^{2} \Delta \omega \frac{1}{\Delta \omega }\right),
\end{eqnarray}
Where $\mathcal{F}_{I}$ represents the part of expression $N_{I}$ that excludes constant factors and the phase-matching function. The transformed form of the integral can be regarded as a integral over different generated modes ($\omega _{s}$, $\omega _{i}$), as well as a summation of the arrangements of pumping components ($\omega _{p}$, $\omega_{p'}$).

Additionally, it is important to note that the expression for the constant $\xi$ differs in definition depending on the coherence. The distinction primarily arises from the influence of coherence on power density, particularly the impact of the spectral line-shape function on power density. The spectrally uncorrelated incoherent field is physically described using discrete variables. In Eq. 2$\&$3, the constant factor is directly given by $\xi _{I} = \frac{1}{2} nc\epsilon _{0} A^{2}$, where the nonlinear effect and spatial mode are expressed by $I(\omega_s, \omega_i) = i\gamma L \text{sinc}(\Delta k L/2) e^{-i\Phi(\omega_s, \omega_i)}$.

For a laser with a continuous spectral distribution, the enhancement due to amplitude coherence is reflected in the power $P_{C}=\frac{1}{2} nc\epsilon _{0} A\left| \int \alpha _{C}( \omega ) d\omega \right| ^{2} \ $, with the effect determined by the spectral line-shape function. This is ultimately expressed as $\Delta\omega_C=\left| \int \alpha _{C}( \omega ) d\omega \right| ^{2} \  $. This expression has the same dimensionality as the bandwidth. It can be understood as the effective bandwidth of coherent light for calculating power density. When calculating the two-photon amplitude, coherence manifests as the coupling between the pump wave $\omega_{p}$ and the overlapping part $\omega_{p'} = \omega _{s} + \omega _{i} - \omega _{p}$. The latter shares the same line-shape, and its amplitude is determined by the self-convolution function. This interaction of coherent pumping can be written as:
\begin{eqnarray}
F'_{C}( \omega _{p'})
&&=\frac{1}{2} nc\epsilon _{0} A\int \alpha_C ( \omega _{p}) d\omega _{p}\nonumber \\ 
&&*A\int \alpha_C ( \omega _{p'}) d\omega _{p'} *[\int d\omega _{p} \alpha_C ( \omega _{p}) \alpha_C ( \omega _{p'})].
\end{eqnarray}
where the two integrals share the same result. To calculate the photon pair generation rate for the orthogonal state $\ket{1_{\omega_{s}}, 1_{\omega_{i}}}$ with continuous distribution, the squared magnitude of Eq. S4 should be divided by $\Delta\omega_C$ to obtain the power density form. The continuous two-photon amplitude constructed in the integral form can be expressed as $F_{C}=\xi_{C}\int d\omega _{p} \alpha_C ( \omega _{p}) \alpha_C ( \omega _{p'})$, where the constant $\xi_{C} = \frac{1}{2} nc\epsilon_0 A^2 \int \alpha_C(\omega) d\omega$.

Besides, when considering the total photon count generated within a channel, under the assumption of a narrow bandwidth, it can be computed by summing over individual contributions. The generation of photon pairs within an asymmetric interval can be described by the following formula:
\begin{eqnarray}
\mathcal{F}_{0}( \Omega ( n_{k}) ,\Omega ( -n_{l})) &&=\Delta \Omega \int _{\frac{n_{k} +n_{l}}{2}}^{\frac{n_{k} +n_{l}}{2} +1} dX^{'}\int _{( n_{k} -n_{l} -1)\frac{\Delta \Omega }{2}}^{( n_{k} -n_{l} +1)\frac{\Delta \Omega }{2}} dY \nonumber \\ 
&& \times exp\left[ -2\left(\frac{Y}{\sigma _{p}}\right)^{2}\right]\mathnormal{\left| J\left( \omega '_{i} ,\omega _{s}^{''}\right)\right| .}
\end{eqnarray}

For $\ k\neq l$, $\mathcal{F}( \Omega ( n_{k}) ,\Omega ( -n_{l})) =0$

\section{Numerical Analysis}

The first procedure is to renormalize the measured parameters. After performing high-pass filtering on the raw data to remove low-intensity noise signals, effective wavelength-intensity(or transmittance) data have transformation into frequency-based information for matching frequency of pump photons with generated photons'. Transformed information with nonlinear distribution accomplish data renormalization with equal interval $d$ via interpolation.

Next, because the signal and idler photons pumped by asymmetry components can't be collected totally, the boundaries of transformed parameters are redefined to facilitate subsequent photons statistics in such situations. Here a function plays a role in the judgement on the maximum and minimum difference of upper and lower bound between spectrum with two collecting channels. The judged values are defined as $D\Omega$ and $d\Omega$ to broaden the range of channels and transmittance of expanded regions is assigned for 0 .In the case of redefined boundaries, it is symmetric about the pump and ensures the integrity of single side count. Additionally, $D\Omega$ and $d\Omega$ which are multiples of $d$ determine a looping variable in procedure 4. 

\begin{figure*}[htbp]
    \centering
    \includegraphics[width=\textwidth]{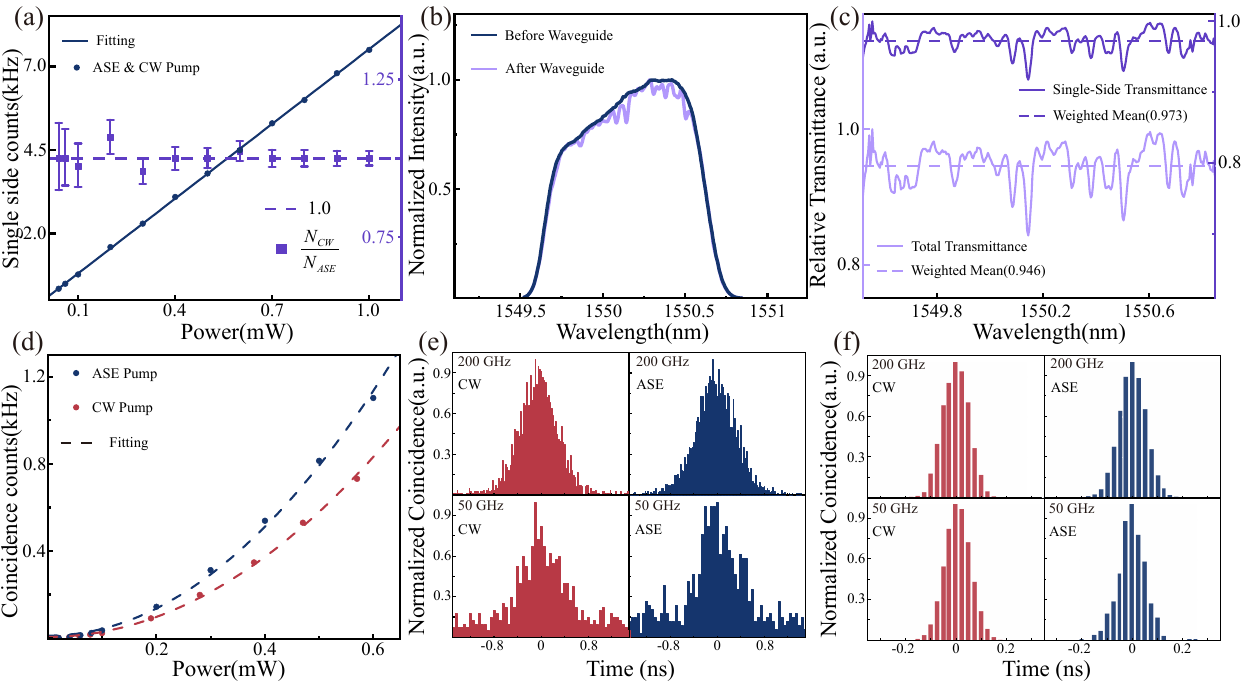}  
    \caption{(a) System noise performance without waveguide insertion; (b) Linear spectral transfer function of the waveguide for a 200GHz ASE; (c) Normalized transmittance as a function of wavelength.
(d) Demonstration of experimental phenomenon reproducibility in different devices ;
(e) Coincidence histogram detected by InGaSn APD;
(f) Coincidence histogram detected by SNSPD.}
    \label{fig:banner}
\end{figure*}

Then, the split-step method is adopted to calculate photon pair generation rate in waveguide. A waveguide of $L$ length is split into $N$ segments of $z=L/N$. The pump passes through the waveguide and the biphoton state generated at each segment are coherently added. Considering the transmission loss, the total amplitude can be iteratively calculated by discretizing the integral forms.

Finally, optimization is indispensable for the multi-variable program that expect for the inevitable loops of pump power and split-step, separated ASE frequency components and broadband photon-pair generation are also increasing the complexity. However, benefiting from the operations above, the processes of interaction within the bandwidth and collection of photons are significantly simplified. The renormalized spectrum can be divided into $M$ segments by unit  length of $d$. Detuning frequency($\delta \Omega$) with the upper bound $D\Omega$ and lower bound $d\Omega$ is defined as the detectable range of photon pairs. We construct two sets of $M*M$ matrices, one represent the information including refractive index and power as a function of pump frequency, the other are related to generated frequency obtained through looping detuning frequency $\delta \Omega$. The algorithm permits the representation for coincidence count, single side count and phase-mismatching by the traversal of detuning frequency, split step and summation of matrices. Pump power, bandwidth and center frequency asymmetry ($\Omega_a$) can be employed as variables in the main loop to explore their impact on photons statistics.

\section{Experimental System Overview}

In this work, as the nanophotonic platform is based on standard silicon nanowires (~450 × 220 nm²), no device fabrication section is included. Although incoherent pumping demonstrated advantages in photon-pair generation rates during the experiment, the difference was not sufficiently significant. To address this, the experimental system has been analyzed in three sections according to their respective functions to exclude potential effects from individual components.

Filtering System: To eliminate sideband noise and isolate the pump light, two sets of cascaded fiber filters were used after EDFA. The devices totally block the noise while maintaining output power stability within ±1\%. After replacing the waveguide with a variable attenuator, the single-side counts for the C20 channel are shown as a function of input power in Figure S1(a). The count rate mainly comes from the linearly increasing Raman noise. When the input power is the same, the ratio of the single-side count rates for both is close to 1. Therefore, it can be concluded that the system does not provide additional selection for light within the filtering bandwidth.

Silicon Nanowire Waveguide: The spectral transfer function of the waveguide plays a crucial role in determining the on-chip power. In this experiment, we employed a packaged waveguide coupled via grating. The frequency-selective property of the grating may influence the coupling efficiency of the CW laser more significantly than the broadband ASE pump. To evaluate the wavelength-dependent transmission characteristics, an ASE pump with a 200 GHz bandwidth was used for comparison.

The normalized spectra and transmittance are shown in Figure S1(b) and S1(c). Since the pump loss is considered only for the single side, the coupling efficiency exhibits some variation across different wavelengths, but the overall difference is not significant. If we assume the maximum single-side transmittance is 1, the weighted average value reaches 0.973. For the 1550.17 nm wavelength used in our experiment, this value is 0.98, with a difference of less than 1\%. Moreover, we demonstrated the same experimental results using a end-face coupled, equal-length silicon nanowire (5dB insert loss) in Figure S1(d).

Detection System: After the photon pairs are separated using DWDM, coincidence counts are recorded with a single-photon detector. In Figure S1(e) and S1(f), we contrast the coincidence histograms with different pump coherences. The photon pairs generated by the two types of pumps have the same spectral distribution.  The value of the FWHM is positively associated with the jitter of the detection and coincidence system, and negatively associated with the bandwidth of the post-filter (DWDM). In the system, the coherence time of generated photon pairs is approximately on the order of 10 ps, so the jitter (>100 ps) determines the bandwidth.

\section{Bandwidth of Incoherent Pumping}
For the temporally incoherent ASE pump, spectral broadening affects both the collection of correlated photon pairs and the state purity. In the main text, the impact of incoherent pumping bandwidth on purity is demonstrated through numerical simulations, showing a positive correlation between bandwidth and purity.

However, a simple consensus is that an increase in bandwidth leads to a decrease in collection efficiency. To quantify this effect, in Figure S2, we numerically 
\begin{figure}[b]
\centering
\includegraphics[width=0.49\textwidth]{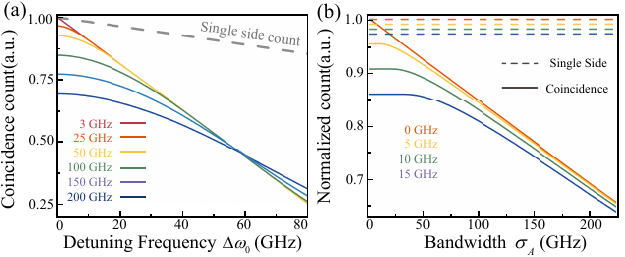}
\caption{\label{fig:epsart}(a) Normalized Coincidence count and single side count versus detuning frequency; (b) Normalized Coincidence count and single side count versus bandwidth.}
\end{figure}
simulate the photon detection efficiency as a function of bandwidth $\sigma_A$ and detuning frequency $\Delta \omega _{0} = \omega _{0} - \omega _{d}$. The pump waveform is idealized as a square wave, and the transmittance of the DWDM after high-pass filtering is set to 1. Figure S2(a) illustrates the impact of detuning frequency on photon pair counting. The line represents the normalized coincidence count, while single-channel counts are shown by the gray dashed line, with differences between them not exceeding 1\%. Therefore, a single representation is used to highlight their sensitivity to detuning frequency.
\begin{figure}[h]
\flushleft
\includegraphics[width=0.49\textwidth]{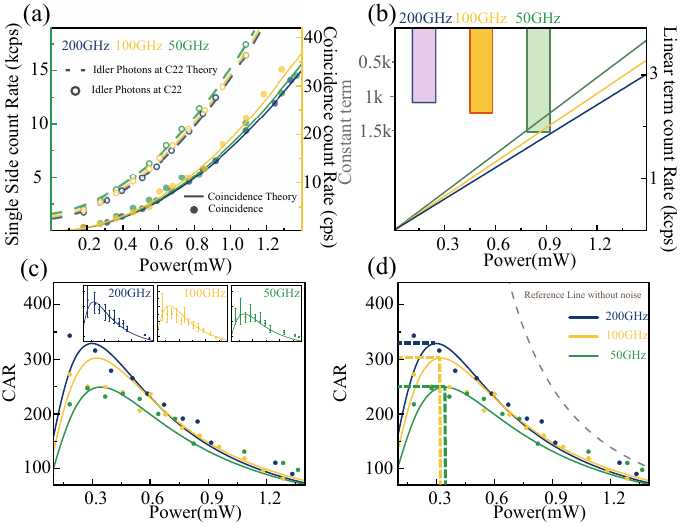}
\caption{\label{fig:epsart}
(a) Coincidence count rate versus pump power and single side count rate of the idler photon from C22 versus pump power with different bandwidth;
(b) Linear component of noise count versus pump power and constant background noise;
(c) Theoretical predictions and experimental CARs versus pump power. The insets are the specific CARs of each bandwidth with error bar.
(d) Theoretical predictions and experimental CARs versus pump power. The insets are the specific CARs of each bandwidth with error bar. The dashed lines correspond to the position of peak CAR.}

\end{figure}
For a broadband pump in SFWM, the interactions of asymmetric components dominate the photon-pair generation process. In this case, some photon pairs cannot be detected simultaneously, leading to a negative correlation between photon count and bandwidth at the initial stages of the function. However, as the detuning frequency increases, more frequency combinations from the broader pump are involved, reducing the relative impact of affected components. This results in a 'resistance effect,' where the function decreases at a slower rate. This behavior is also shown in Figure S2(b). As bandwidth increases, the overall trend of the function is a decrease. For single side count, detuning frequency alters the phase-matching function, reducing efficiency, while the influence of bandwidth becomes negligible.

In the experiment, the influence of bandwidth for incoherent correlation properties is represented in Figure S3(a)-(c). As predicted in Figure S2, there are no obvious differences in idler photon count. The dashed lines illustrated in Figure S3(a) maintain a similar quadratic growth trend, confirming this conclusion, and numerically calculated curves are also in great consistency with results. In this case, noise count is the main contributor to the bandwidth performance. For ease of comparison, in Figure S3(b) we depict linear terms and constant terms of idler count respectively. The request for narrower ASE pumped bandwidth at target pump power should be satisfied by higher amplification of seed, which leads to an increase of background noise and enhancement of linear effect in sidebands. Affected by this, the value and position of the peak CAR are also changed as shown in Figure S3(c), see more details in section "CAR Curve".

\begin{figure*}
    \centering
    \includegraphics[width=\textwidth]{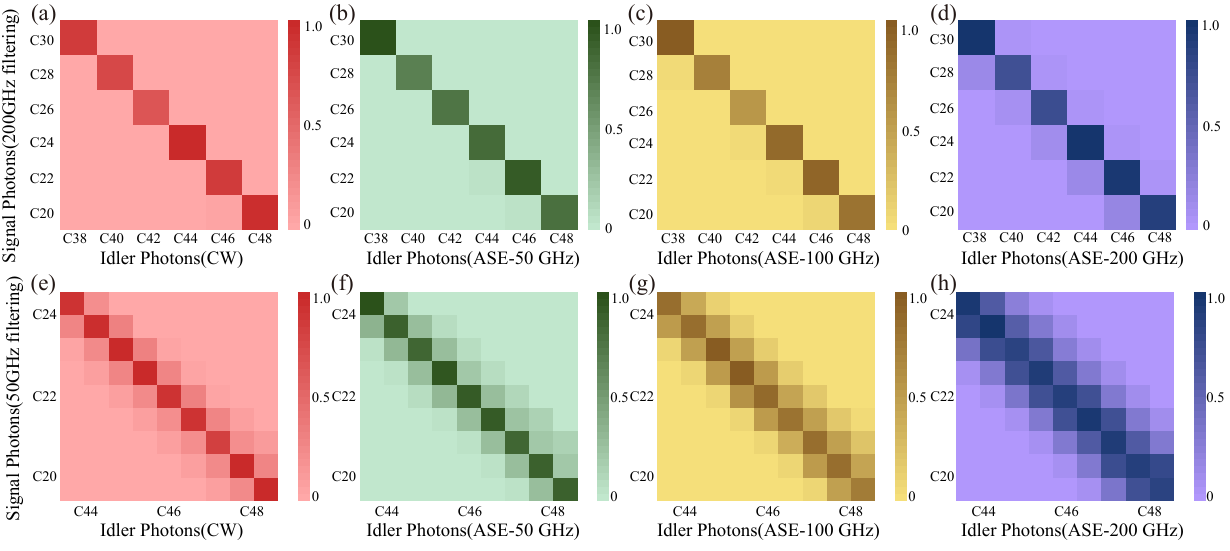}  
    \caption{Joint spectral Intensity. (a)-(d) 200 GHZ DWDM; (e)-(h) 50 GHZ DWDM.}
    \label{fig:banner}
\end{figure*}

In order to study the distribution of correlated photon pairs, we measured the joint spectral intensity using two types of DWDMs and pumps with four bandwidths in Figure S4. Due to the broad bandwidth and center frequency detuning, as well as the overlap between adjacent channels, coincident events can also be detected in the asymmetric channels. The results with the 50 GHz DWDM more clearly show the distribution of photon pairs generated by pumps with varying bandwidths.

\section{CAR Curve}
Generally, CAR curve is depicted by fitting the measured data. In this study, we can carry on a detailed analysis of the curve's trend due to the theoretically calculated results and expressions. The expanding expression of CAR can be written as:
\begin{equation}
CAR = \frac{B_{cc}\cdot P^2}{\Delta \tau(B_{sc1}\cdot P^2+a_{1}\cdot P+N_{1}) \cdot (B_{sc2}\cdot P^2+a_{2}\cdot P+N_{2})},
\end{equation}
where $B$ refers to the brightness for the generated photon-pair. $a$ and $N$ refer to intensity of noise count. Without them, CAR function will show a monotonically decreasing trend as shown by the gray line in Figure S3(d). Subscript is on behalf of count type and channel. For verifying the effect of each component on the CAR extremum, we calculate its first-order derivative of power. When the pump power satisfies the equation
\begin{eqnarray}
(a_{1}\cdot N_{2}+a_{2}\cdot N_1)\cdot P+2N_{1}N_{2}&&=2B_{sc1}B_{sc2}\cdot P^4\\
&&+(a_{1}\cdot B_{sc2}+a_{2}\cdot B_{sc1})\cdot P^3 \nonumber.
\end{eqnarray}

Although the solution of the equation can be solved using the formula of the fourth degree equation, it's unworthy for the research on the impact of various parameters. Both sides of the equation can be regarded as a slowly growing linear noise function $f_{noise}$ and a rapidly growing nonlinear coincidence function $f_{cc}$ respectively. Assuming that $P=P_{1}$ is the solution of Eq. S8 corresponding to pumped by ASE with broad bandwidth. The parameters are related to the proportion of sideband photons in the noise. In our system, the relationship of them can be described as $1<\mu_1<\mu_2$. Additionally, the influence of bandwidth is negligible impact on single side count from SFWM. Based on conditions above, we can construct function $\Delta=f_{noise}-f_{cc}$ to illustrate the change of extremum. It can be expressed as

\begin{eqnarray}
\Delta(P_1;\mu_1 a_{1,2};\mu_2 N_{1,2})&&=(\mu_1 \mu_2-1)f_{cc}(P_1;a_{1,2};N_{1,2})\\
&&+2(\mu_2^2-\mu_1 \mu_2)N_{1}N_{2}>0, \nonumber
\end{eqnarray}
where $P=P_{1}$, $a$ and $N$ are replaced by $\mu_1 a$ and $\mu_2 N$ to represent noise photons introduced by narrower bandwidth. The result shows that higher pump power is required to reach up to extremum position, and in that case the maximum CAR will be lower for monotonically decreasing trend without noise count.

\section{Spectrally Multiplexed Correlation}

\begin{figure}[b]
\flushleft
\includegraphics[width=0.49\textwidth]{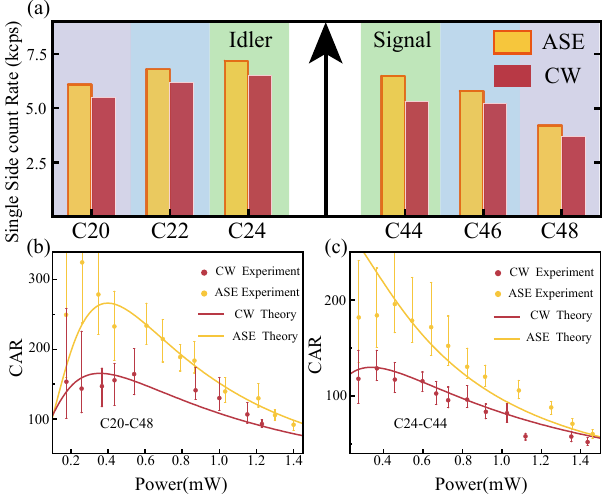}
\caption{\label{fig:epsart}
(a) Spectra of the correlated signal and idler photons;
(b) Theoretical plot and experimental CARs versus pump power from C20-C48;
(c) Theoretical plot and experimental CARs versus pump power from C24-C44.}
\end{figure}

The correlated photon pairs generated via the SFWM process exhibit inherent spectral multimode characteristics. In the experiment, we investigate the generation of multiplexed photon pairs. Despite limitations imposed by the range of DWDM channels and the bandwidth of post-filters, quantum correlation properties between channels C20–C48 and C24–C44 are successfully characterized. Theoretically predicted and experimentally measured CAR values show strong agreement. Figure S5(a) illustrates single-side counts under 0.6 mW pumping with a 100 GHz ASE source and a CW laser. The amplification of wavelength-paired resonances and the reduction in channel intervals between signal and idler photons reflect reduced phase matching. Compared to SPDC, the convolution effects of the field and medium in SFWM are significantly weaker due to the slowly varying phase-matching function.

\


